\documentclass[graphicx,times]{mn2e}
\usepackage{epsfig}
  
\newcommand{\gsim}{\mathrel{\hbox{\rlap{\lower.55ex \hbox {$\sim$}}
                   \kern-.3em \raise.4ex \hbox{$>$}}}}

\title[Young multiple stars]{On the properties of young multiple stars}

\author[Delgado-Donate, Clarke, Bate \& Hodgkin]{E. J. Delgado-Donate$^{1,2}\,$\thanks{E-mail:
edelgado@astro.su.se}, C. J. Clarke$^1$, M. R. Bate$^3$, S. T. Hodgkin$^1$\\
$^1$Institute of Astronomy, University of Cambridge, Madingley Road,
Cambridge, CB3 0HA\\
$^2$Stockholm Observatory, AlbaNova University Centre, 106 91 Stockholm, Sweden\\
$^3$School of Physics, University of Exeter, Stocker Road, Exeter EX4
4QL}

\begin{document}

\date{}

\pagerange{\pageref{firstpage}--\pageref{lastpage}} \pubyear{2003}

\maketitle

\label{firstpage}

\begin{abstract}
We present numerical results on the properties of young binary and
multiple stellar systems. Our analysis is based on a series of SPH $+$
$N$-body simulations of the fragmentation of small molecular clouds,
that fully resolve the opacity limit for fragmentation. These
simulations demonstrate that multiple star formation is a major
channel for star formation in turbulent flows. We have produced a
statistically significant number of stable multiple systems, with
components separations in the range $\sim 1-10^3$ AU. At the end of
the hydrodynamic stage (0.5 Myr) we find that $\approx 60\%$ of stars
and brown dwarfs are members of multiples systems, with about a third
of these being low mass, weakly bound outliers in wide eccentric
orbits. Our results imply that in the stellar regime most stars are in
multiples ($\approx 80\%$) and that this fraction is an increasing
function of primary mass. After $N$-body integration to 10.5 Myr, the
percentage of bound objects has dropped to about 40\%, this decrease
arising mostly from very low mass stars and brown dwarfs that have
been released into the field. Brown dwarfs are never found to be very
close companions to stars (the {\it brown dwarf desert at very small
separations}), but one case exists of a brown dwarf companion at
intermediate separations (10 AU). Our simulations can accommodate the
existence of brown dwarf companions at large separations, but only if
the primaries of these systems are themselves multiples.

We have compared the outcome of our simulations with the properties of
real stellar systems as deduced from the IR colour-magnitude diagram
of the Praesepe cluster and from spectroscopic and high-resolution
imaging surveys of young clusters and the field. We find that the
spread of the observed main sequence of Praesepe in the
$0.4-1$M$_\odot$ range appears to require that stars are indeed
commonly assembled into high-order multiple systems. Similarly,
observational results from Taurus and $\rho$ Ophiuchus, or moving
groups such a TW~Hydrae and MBM~12, suggest that companion frequencies
in young systems can be indeed as high as we predict. The comparison
with observational data also illustrates two problems with the
simulation results. Firstly, low mass ratio ($q < 0.2$) binaries are
not produced by our models, in conflict with both the Praesepe colour
magnitude diagram and independent evidence from field binary
surveys. Secondly, very low mass stars and brown dwarf binaries appear
to be considerably under-produced by our simulations.
\end{abstract}


\section{Introduction}

The formation of binary and multiple systems provides one of the most
exacting areas in which star formation theory can be compared with
observational data. Stars are known to have a binary frequency in
excess of 50\%, both in the field (Duquennoy \& Mayor 1991, henceforth
DM91; Fisher \& Marcy 1992; Halbwachs et al. 2003) and in clusters
(K\"ohler \& Leinert 1998; Patience et al. 1998; Mathieu et al. 2000;
Bouvier et al. 2001).  For pre-main-sequence stars this frequency
seems to be even higher (Reipurth \& Zinnecker 1993; Simon et
al. 1995; Duch\^ene 1999; Reipurth 2000). Thus, understanding the
formation of multiple stars becomes necessary if we are to understand
star formation in general. Recently, Bate, Bonnell \& Bromm (2003;
henceforth BBB) have shown that multiple stars are a natural byproduct
of the collapse and fragmentation of turbulent molecular clouds. In
particular, close binary stars can be indirectly formed out of the
fragmentation of a molecular cloud core by means of dynamical
interactions with other stellar objects in the vicinity, as well as
interactions with circumbinary discs (Bate, Bonnell \& Bromm
2002b). However, this sort of calculation (that fully resolve
fragmentation in a cloud of e.g. 50 M$_\odot$ of gas in BBB's case) is
very demanding computationally, and a wide range of initial conditions
cannot easily be explored. In addition, such calculations remain
inconclusive about the properties of multiple stars (in particular the
existence of long lived multiples and the production of wide binaries)
since the evolution of the resulting stellar systems is not followed
until decay to a stable hierarchical configuration is achieved. For
example, out of the 50 objects formed in the BBB simulation, 18 are
contained in non-hierarchical groupings containing 11 and 7
members. Since, in this simulation, only the more compact groupings
have had time to decay to a stable configuration, BBB understandably
focused their attention on the implications of their work for {\it
close} binary formation.

In this paper we have taken an alternative approach which is a natural
complement to the BBB simulations. Through the modelling of less
massive clouds over longer timescales, we have been able to follow
wide binary formation and the production of long lived multiple
systems.  We find that by modelling an ensemble of isolated cores of
mass 5~M$_\odot$, we can improve the number of stars formed per CPU
hour by a factor 7 compared with the BBB calculation, which comprised
a total mass of 50~M$_\odot$.  This economy stems from the fact that
by focusing on individual dense cores, we dispense with the
computational expense of following the diffuse gas in the BBB
simulation. We remark, however, as a caveat, that we find that the
formation of multiples proceeds {\it hierarchically}, with structures
on any scale being progressively modified by interactions on larger
scales. Thus in these models with scale free turbulence, {\it any}
upper mass cut-off (i.e. the finite mass of the system modelled) may
have some influence on the multiple systems produced. By comparing our
results with those of BBB, we are able to assess how sensitive our
results are to the total mass of the region simulated.

The computational economies that we gain through the simulation of
less massive cores, allows us to instead concentrate our efforts on
the longer timescale integration of the multiple systems produced.  We
wish to understand how our results are affected if we {\it throw away}
the gas component at a certain evolutionary stage (when 60\% of the
initial gas has been accreted) and thereafter evolve the system as an
$N$-body ensemble. We follow the stellar dynamical $N$-body problem
for 10~Myr (this typically corresponding to $10^4$ orbits of multiple
systems at the median separation), using the {\sc nbody1} code, by
Aarseth (Aarseth 1963). Another advantage that we reap from this less
computationally expensive approach is that we gain improved statistics
(i.e. we form $\approx 150$ stars and brown dwarfs, compared with the
$\approx 50$ of BBB), and from this dataset are able to extract
reasonable statistics, in a manner similar to Sterzik \& Durisen
(1998; 2003) for the purely $N$-body case. We stress that our
simulations share the property of the BBB simulations that they
resolve the opacity limit for fragmentation (Low \& Lynden-Bell 1976;
Rees 1976) and that, assuming that fragmentation does not occur at
densities greater than those at which the gas becomes opaque to
infrared radiation, these calculations are able to model the formation
of all the stars and brown dwarfs that, under the initial conditions
imposed, can be produced. Our spatial resolution limit for binaries
allows us to study a wide range of separations, and the particle
numbers we employ allow us to model accretion discs around the
protostars which are as long lived as those modelled by BBB.

Our study has four major findings:
\begin{itemize}
\item The incidence of multiple systems (with N $>$ 2) is {\it high}
(8 out of 18 N $>$ 1 systems at $t=10.5$ Myr) and we therefore test
whether our results are compatible with the photometric width of the
main sequence in young clusters. The companion frequency varies
significantly during the pure $N$-body evolution of the systems.
\item Although we produce a variety of triples, quadruples and higher
order multiples, they tend to follow a characteristic pattern of {\it
internal mass distribution}. This involves each binary pair having a
mass ratio that does not deviate strongly from unity, and likewise,
for quadruples, the mass ratio of the total mass in each binary is not
greatly different from unity (i.e. mass ratios in the range $0.5-1$
are favoured in each case). The exception to this is that multiple
systems are commonly orbited by {\it low mass outliers} on eccentric
orbits, which are the result of incomplete ejections of low mass
objects from multiple systems. We discuss how deep images of the
environs of multiple systems, and spectroscopic studies of the
primaries of extreme mass ratio binaries, may be used to test this
prediction.
\item The multiplicity fraction is an increasing function
of primary mass. The brown dwarf desert at {\it very small
separations} is reproduced by our models.
\item We confirm the result of previous simulations that brown
dwarf binaries indeed appear to be under-produced by turbulent
fragmentation calculations.
\end{itemize}

The structure of this paper is as follows. In Section 2 the
computational method and initial conditions applied to our models are
described. The results on multiple stars are given in Section 3. In
Section 4 we perform a detailed comparison of our results with
available observational data, and suggest future
experiments. Our conclusions are given in Section 5.

\section{Computational method}

 The results presented in this paper were obtained from the same
simulations introduced in Delgado-Donate, Clarke \& Bate (2004;
henceforth DCB04). A detailed description of the SPH code that was
used to perform those calculations as well as the initial conditions
imposed is given there. Here we briefly summarise the computational
method.

The calculations were performed using a 3D hybrid SPH $N$-body code,
with variable smoothing lengths. The SPH equations are solved using a
second order Runge-Kutta-Fehlberg integrator with individual timesteps
for each particle (Bate, Bonnell \& Price 1995). We use the standard
form of artificial viscosity (Monaghan \& Gingold 1983) with strength
parameters $\alpha_{\rm v} = 1$ and $\beta_{\rm v} = 2$.

\subsection{Equation of state}

 We have assumed that the gas becomes optically thick when the density
$\rho$ reaches a critical value $\rho_{\rm c} = 10^{-13} {\rm
g~cm^{-3}}$. This density defines the so-called opacity limit for
fragmentation, which sets a minimum fragment mass of a few Jupiter
masses (Low \& Lynden-Bell 1976; Boss 1988; BBB).

To model the opacity limit for fragmentation without performing full
radiative transfer, we use an equation of state given by $p$ = $K$
$\rho^{\eta}$, where $p$ is the pressure and $K$ is a measure of the
entropy of the gas. The value of $\eta$ changes with density as:
\begin{equation}
\eta = \cases{\begin{array}{rl}
1, & \rho \leq 10^{-13}~ {\rm g~cm}^{-3}, \cr
7/5, & \rho > 10^{-13}~ {\rm g~cm}^{-3}. \cr
\end{array}}
\end{equation}
The gas is assumed to consist of pure molecular hydrogen ($\mu$ = 2),
and the value of $K$ is such that when the gas is isothermal, $K$ =
$c_{\rm s}^2$, with the sound speed $c_{\rm s} = 1.85 \times 10^4 {\rm
cm s}^{-1}$ at $T = 10$ K. The pressure is continuous when the value
of $\eta$ changes.

\subsection{Sink particles}

 A sink particle is inserted when the central density of a fragment
exceeds $\rho_{\rm s} = 10^{-10} {\rm g~cm}^{-3}$, well above the
critical density $\rho_{\rm c}$. Sink particles are point masses with
an accretion radius, so that any gas particle that falls into it and
is bound to the point mass is accreted. In the present calculations,
the accretion radius $R_{\rm sink}$ is constant and equal to 5
AU. Therefore, discs around sink particles will be resolved only if
their radii $\gsim 10$ AU. Sink particles interact with the gas only
via gravity and accretion. 

The gravitational acceleration between two sink particles is Newtonian
for $r \geq 4$ AU, but is smoothed within this radius using spline
softening (Benz 1990). The maximum acceleration occurs at $r \sim 1$
AU; therefore, this is the minimum binary separation that can be
resolved.

\subsection{Initial Conditions}

 We have performed 10 different calculations of the fragmentation of a
small-scale, turbulent molecular cloud, each of them under almost
exactly the same initial conditions. Each cloud core is spherical, has
a mass of 5 M$_\odot$, a radius of $\approx 10^4$ AU, and an initial
uniform density of $10^{-18} {\rm g~cm}^{-3}$. At the initial
temperature of 10 K, the mean thermal Jeans mass is 0.5 M$_\odot$,
i.e. the Jeans number of the cloud is 10. The global free-fall time of
the cloud $t_{\rm ff}$ is $\approx 10^{5}$ yr.

We have imposed an initial supersonic turbulent velocity field on the
gas, in the same manner as Ostriker, Stone \& Gammie (2001) and
BBB. We generate a divergence-free random Gaussian velocity field with
a power spectrum $P(k) \propto k^{\alpha}$, where $k$ is the
wavenumber and $\alpha$ is the power index, {\it which we have set to
$-3$ in half of the simulations and $-5$ in the other half}. The
velocity field is normalised so that initially it is in equipartition
with the gravitational potential energy of the cloud core.

 These simulations do not include magnetic fields, as we have tried to
isolate a particular hydrodynamical fragmentation problem to
characterise the properties of the resulting stellar systems. We have
not included in our models any mechanical or radiative feedback
mechanism. This may be an appropriate choice, since the maximum
stellar mass in these simulations does not exceed 1 M$_\odot$, whereas
the most powerful winds and photoionisation fronts in star-forming
regions are produced by much more massive stars.

\subsection{Resolution}

 The local Jeans mass must be resolved throughout the calculation
(Bate \& Burkert 1997; Truelove et al. 1997; Whitworth 1998),
otherwise some of the fragmentation might be artificially enhanced or
suppressed. In order to model a 5 M$_\odot$ cloud core with critical
density $\rho_{\rm c}$ we need to use $3.5 \times 10^5$ particles.

Each of the hydrodynamic calculations that are discussed in this paper
required $\approx 4000$ CPU hours on the SGI Origin 3800 Computer of
the United Kingdom Astrophysical Fluids Facility (UKAFF).

\begin{table*}
\begin{minipage}{175mm}
\caption{Multiple systems configurations. Each main row shows the
results obtained for a given simulation: the first 5 correspond to the
$\alpha$ = $-$ 3 case, and the following 5 to the $\alpha$ = $-$ 3
case. The last row shows the results for all the simulations
combined. For each simulation, every column except the first two are
divided in two rows: the upper one refers to the results at $t = 0.5$
Myr, i.e., at the end of the hydrodynamic calculation; the bottom row
correspond to the results at $t = 10.5$ Myr, i.e., at the end of the
$N$-body integration. The {\it Configuration} column provides a
symbolic representation of the hierarchical distribution of the stars
and brown dwarfs making up a given multiple. The numbers can be
interpreted as arbitrary flags assigned to each of the components of a
given multiple (in fact, object 1 is simply the first star to have
formed in that calculation, object 2 the second and so forth). Two
numbers linked by a dash and between parentheses represent a binary
system. Subsequent dashes denote further bonds with other objects; and
square and curly brackets enclose those objects that belong to
sub-systems with higher order in the hierarchy. For example, in
simulation $\alpha$3A there is one multiple system which consists of a
binary (3-5), orbiting a single (dubbed 0), thus making the triple
[0-(3-5)]. This triple system is bound to another binary (1-2), thus
forming a quintuple system \{[0-(3-5)]-(1-2)\}. Additional objects
enclosed in parentheses represent {\it outliers}: individual very low
mass stars or brown dwarfs also bound to the multiple but moving in
eccentric orbits at large separation from the CoM of the system. A
plus sign is used to highlight that a given simulation produced two or
three mutually unbound multiples.}
\begin{tabular}{lcc@{;}clcl}
\hline \hline 
Simulation & N$_{\rm tot}^{a}$ & \multicolumn{2}{c}{[N$_{\rm m}$ ;
N$_{\rm s}$]$^{b}$} & Configuration$^{c}$ & N$_{\rm out}^{d}$ &
Sep$_{\rm out}^{e}$ [AU] \vspace{0.5mm} \\
\hline
$\alpha$3A$^f$ & 16 & 9     & 7 & 
$((((\{[0-(3-5)]-(1-2)\}-14)-15)-4)-6)$ & 4 & $10^4$ \\
     ~     & ~  & 6     & 10&
$(\{[0-(3-5)]-(1-2)\}-15)$      &   1  & $4 \times 10^3$\\ 
\hline
$\alpha$3B & 21 & 8     & 13 & 
$(((\{[(1-3)-7]-(0-2)\}-18)-6)-10)$ & 3 & $10^3$ \\
     ~     & ~  & 5     & 16 &
$\{[(1-3)-7]-(0-2)\}$           &   0  &  $10^3$\\
\hline
$\alpha$3C & 17 & 11    & 6  &
$(((([\{[(0-2)-1]-(3-4)\}-(6-7)]-10)-13)-12)-5)$ & 4 & $10^3$\\
     ~     & ~  & (3,2,2)     & 10 &
$[(0-2)-1]$ ~ + ~ $(3-4)$ ~ + ~ $(6-7)$ & 0 & $10^2$\\
\hline
$\alpha$3D & 14 & 10    & 4  &
$((((\{[(0-1)-(11-10)]-(3-2)\}-6)-12)-8)-13)$ & 4 & $5 \times
10^3$ \\
    ~      & ~  & (3,2) & 9 &
$[(0-1)-6]$ ~ + ~ $(3-10)$        & 1 & 10 \\
\hline
$\alpha$3E & 17 & 9     & 8  & 
$(((\{[(1-2)-(7-13)]-(0-4)\}-16)-5)-11)$ & 3 & $7 \times 10^4$\\
     ~     & ~  & 6     & 11 &
$\{[(1-2)-(7-13)]-(0-4)\}$    & 0 & $7 \times 10^3$\\
\hline \hline
$\alpha$5A & 12 & (5,2) & 5  & 
$([(0-1)-(7-2)]-4)$ ~ + ~ $(8-9)$ & 1 & $2 \times 10^3$ \\
    ~      & ~  & (4,2) & 6  &
$[(0-1)-(7-2)]$ ~ + ~ $(8-9)$ & 0 & $10^2$\\
\hline
$\alpha$5B & 7  & 6     & 1  & 
$(([(1-2)-(0-4)]-5)-6)$ & 2 & $2 \times 10^3$ \\
     ~     & ~  & (2,2) & 3  &
$(1-2)$ ~ + ~ $(0-4)$     & 0 & 10 \\
\hline
$\alpha$5C & 11 & 5     & 6  & 
$(\{[(0-5)-2]-3\}-9)$ & 2 & $10^5$\\
    ~      & ~  & 4     & 7  &
$([(0-5)-2]-9)$         & 1 & $3 \times 10^4$\\  
\hline
$\alpha$5D & 16 & 9     & 7  & 
$(((\{[(4-0)-5]-[(1-9)-14]\}-15)-11)-8)$ & 5 & $10^4$\\
     ~     & ~  & (2,2) & 12 &
$(1-9)$ ~ + ~ $(4-0)$     & 0 & 10 \\
\hline
$\alpha$5E & 14 & (6,2,4) & 2  & 
$(\{[(5-6)-(4-8)]-12\}-9)$  ~ + ~ $(13-10)$ ~ + ~ $[(1-2)-(0-3)]$ &  2 & 
$4 \times 10^3$\\
    ~      &  ~ & (4,2,2) & 6  &
$[(1-2)-(0-3)]$ ~ + ~ $(5-6)$ ~ + ~ $(13-10)$   & 0 & $2 \times 10^2$\\
\hline \hline
All        & 145&  86     & 59 & 
\hspace{2cm} $[ ~ 40\%~ , ~20\% ~ ]$ ~ + ~ 40\%  & 30 & \hspace{0.5cm}
--\\
      ~    &  ~ &  55     & 90 &
\hspace{2cm} $[ ~ 36\%~ , ~~~2\% ~ ]$ ~ + ~ 62\%  & 3 & \hspace{0.5cm}
--\\
\hline \hline  
\end{tabular}
\medskip
\\
$^{a}$ N$_{\rm tot}$: total number of stars and brown dwarfs formed
in a given calculation.\\
$^{b}$ N$_{\rm m}$: number of stars/brown dwarfs bound to a given
multiple system; N$_{\rm s}$: number of singles.\\
$^{c}$ In the last row {\it Configuration} gives: \% of
inner companions , \% of outer companions (as defined in $^{d}$ below)
+ \% of unbound objects\\
$^{d}$ N$_{\rm out}$: number of bound objects with high eccentricities
and separations larger than 1000 AU from the innermost component of
the multiple system.\\
$^{e}$ Sep$_{\rm out}$: Separation between innermost and outermost
components of the multiple, in AU.\\
$^{f}$ The [0-(3-5)] triple remains unstable at 10.5 Myr.\\
\end{minipage}
\end{table*}

\subsection{$N$-body calculations}

The hydrodynamic evolution of each cloud is followed until 60\% of the
initial gas particles are accreted. At this point, the remaining gas
is removed and thereafter the system is evolved as an $N$-body
ensemble for 10 Myr, using the numerical code {\sc nbody1}, by Aarseth
(Aarseth 1963; see Aarseth 1999 for an updated description of the
code). {\sc nbody1} is a simple $N$-body algorithm which computes the
gravitational forces between particles by direct summation. The
numerical scheme is based upon the expansion of the gravitational
force in a fourth-order polynomial with divided
differences. Individual timesteps are also implemented. The Newtonian
potential is softened in a manner that mimics a Plummer sphere.

{\sc nbody1} lacks any regularization scheme for the treatment of
close binaries, in contrast with subsequent more sophisticated {\sc
nbody} codes. Close binaries can nevertheless be accurately integrated
provided that a sufficiently small softening length and $\eta$
parameter (which controls the size of the timesteps) is chosen. We
took softening lengths of $0.1 \times$ the softening length of the SPH
simulations, and $\eta$ parameters smaller than $10^{-2}$, and found
that energy and angular momentum were conserved to an accuracy better
than one part in one million. Each $N$-body simulation required only
$\approx 50$ CPU hours to be completed.

\section{The properties of the multiple stars}

The hydrodynamical evolution of the cloud produces shocks which
decrease the turbulent kinetic energy that initially supported the
cloud. In parts of the system, gravity begins to dominate and dense
self-gravitating cores form and collapse. These dense cores are the
sites where the formation of stars and brown dwarfs occurs. The
turbulence decays on the dynamical timescale of the cloud (as found by
Mac Low et al. 1998; Stone, Ostriker \& Gammie 1998; and BBB, among
others), and star formation begins just after 1 to 1.5 global
free-fall times $t_{\rm ff}$. As mentioned before, the hydrodynamical
calculations were stopped when 60\% of the gas has been accreted. In
terms of $t_{\rm ff}$ this means that, for the $\alpha$ = $-3$ and
$\alpha$ = $-5$ calculations (henceforth $\alpha3$ and $\alpha5$), we
follow the evolution of the cloud for $\approx 4 t_{\rm ff}$ and $5.5
t_{\rm ff}$, respectively (i.e. an average of $\approx 0.5$
Myr). Altogether, the calculations produce 145 stars and brown
dwarfs. An analysis of the dependence of the statistical properties of
the resulting stars and brown dwarfs (e.g. the IMF) on the different
initial conditions imposed is presented in DCB04. Here, we focus on
the formation of multiple stellar systems and their internal structure
(internal distribution of masses and separations).

\subsection{Internal structure}

Both sets of initial conditions result in the formation of a large
number of multiple stars. The statistical properties of the multiple
systems are not very sensitive to the slope of the initial turbulent
power spectrum (DCB04),  implying that they are more sensitive to
the dynamical and competitive accretion processes within each
mini-cluster rather than the large scale morphology of the
cloud. Henceforth we analyse the combined dataset for the $\alpha3$
and $\alpha5$ runs.

Independently of whether the gas is initially dominated by large-
($\alpha5$) or small- ($\alpha3$) scale turbulent motions, a series of
localised, dense pockets of gas are formed within the cloud. A
pressure-supported object forms within each of these dense clumps,
first in isolation but soon surrounded by an accretion disc. Initially
the mass of the disc is comparable to, and often greater than the mass
of the central object. Thus, the disc is prone to the appearance of
gravitational instabilities which, in most cases, result in the
fragmentation of the disc into one or more protostellar objects
(Bonnell 1994; Bonnell \& Bate 1994; Burkert, Bate \& Bodenheimer
1997; Whitworth et al. 1995). The formation of this first star
generally occurs in the lowest of the local potential
minima. Surrounding condensations with slightly lower gas densities
form additional stars (e.g. in the filaments whose intersection
generated the first dense clump). Both the stars and the residual gas
are attracted by their mutual gravitational forces and fall towards
each other. The interactions between the gas and the protostars
dissipate some of the kinetic energy of the latter (Bonnell et
al. 1997), allowing the stellar objects to rapidly come close to the
initial star and its disc-born companions to form a high-density
sub-cluster containing from 2 up to 8 stars: a small-$N$ cluster. This
process repeats itself in other parts of the cloud. Given the size of
the cloud we have modelled (and consequently, the number of Jeans
masses initially present in the system), no more than 3 of these
star-forming sites are ever produced. Subsequently, sub-clusters are
attracted to each other and merge to form the final
mini-cluster. Thus, the star formation process is hierarchical in
nature, as has been vividly illustrated (for a 1000 M$_\odot$ cloud)
by Bonnell, Bate \& Vine (2003).

 These sub-clusters bear much similarity with the small-$N$
clusters modelled by Delgado-Donate, Clarke \& Bate (2003; henceforth
DCB03) in the sense that, initially, the cluster components are
arranged in a dynamically unstable configuration. As the systems seeks
to attain stability, the cluster breaks up: the most massive
components form a hierarchical multiple (typically a binary or a
triple system) whilst the low mass components are ejected, either to
large separations or from the cloud altogether. In contrast with the
DCB03 simulations, the number of stellar seeds is not fixed and
therefore further star formation events can take place, introducing
dynamical instability again in the system. Sub-cluster merging
provides an additional complication to the simple small-$N$ cluster
picture, as a multiple system can be driven to the proximity of
another one so that further interactions take place.

\begin{figure*}
\begin{center}
\caption{Mass ratios and semi-major axis (at 0.5 Myr). Left panel:
mass ratio $q$ vs primary mass (in M$_\odot$), for all the systems
formed in the simulations. Diamonds correspond to binaries (either
independent or bound to larger structures), triangles denote triples
(same as for binaries), squares quadruples (same), asterisks
quintuples (same) and crosses higher-order multiples. The values of
$q$ must be interpreted as meaning the ratio of the mass of the
outermost object of the sub-system under consideration to the total
mass of all the objects interior to it (i.e. for a triple [(1-2)-3],
$q$ = m3/(m1+m2)). Primary mass refers to the total mass of all the
objects interior to that for which the mass ratio is being
calculated. Between the solid and the dashed line are located those
systems with a brown dwarf companion. Brown dwarf binaries would
appear to the left of the dashed line. Right panel: semi-major axis
(in AU) vs primary mass (in M$_\odot$) for all the multiple systems
produced by the simulations. Symbols as in left panel.}
\centerline{\epsfig{file=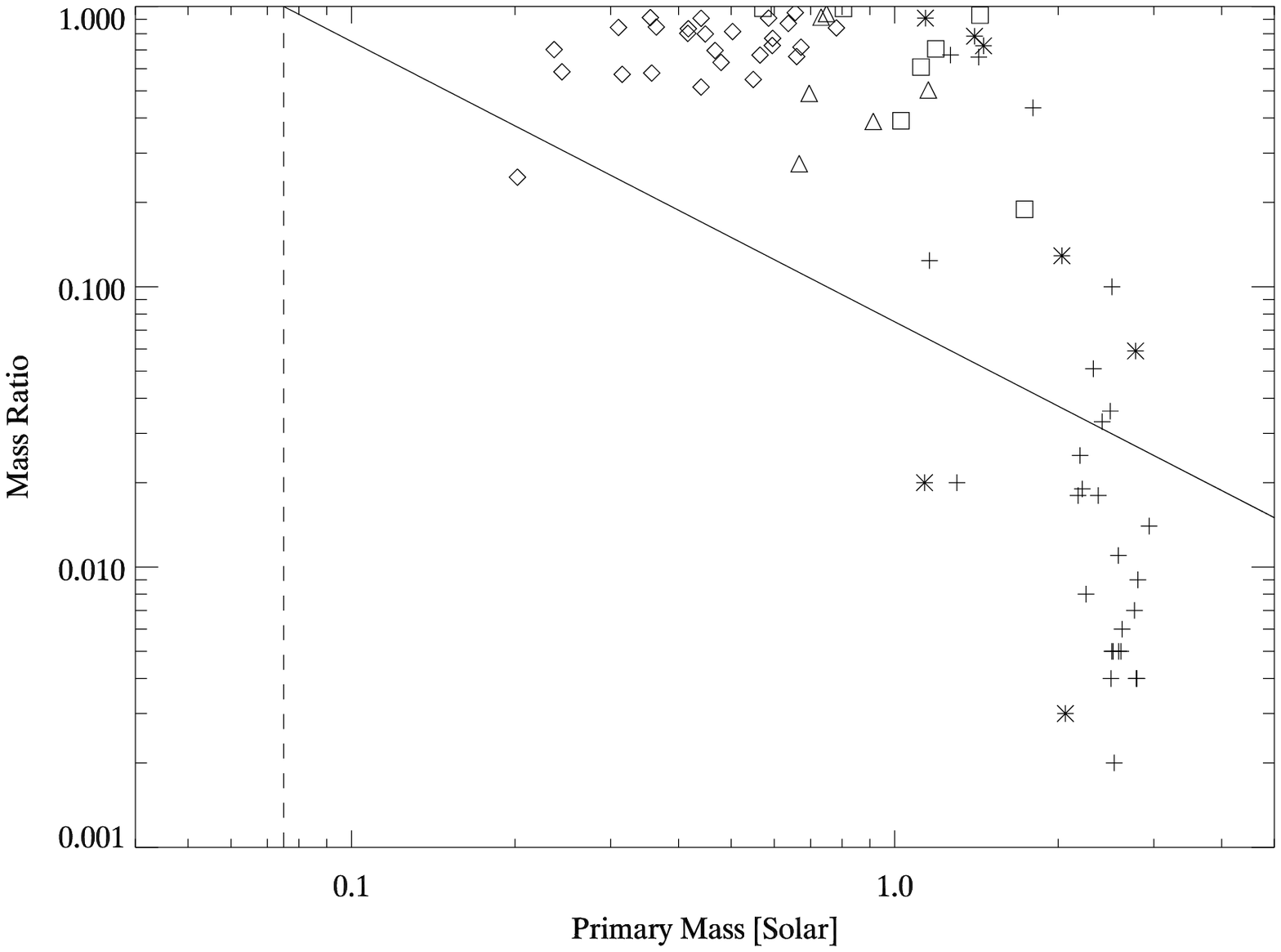,height=6.5cm}\epsfig{file=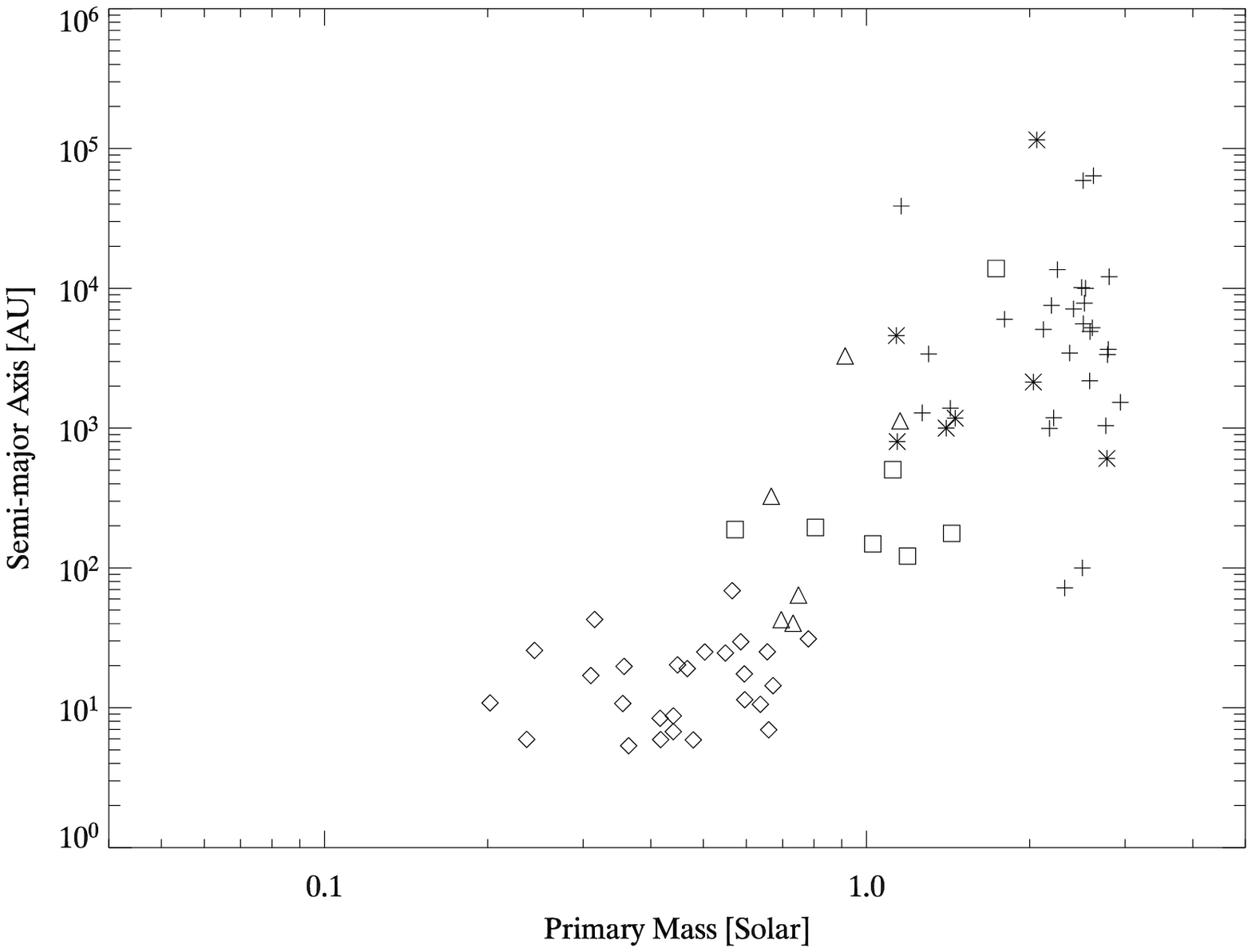,height=6.5cm}}
\end{center}
\end{figure*}

The fact that low mass components are the prime candidates to be
ejected means that, given the high stellar density of each sub-cluster
and the tendency of sub-clusters to interact with each other, few
bound pairs involving a low mass component -- i.e. low mass ratio,
wide or low mass pairs -- can survive the interaction with other
cluster members. The binding energy of these pairs is too low. In
addition, in the present simulations accretion discs surround most of
the stellar objects formed, and when surrounding a binary, tend to
drive the mass ratio $q$ to high values (Bate \& Bonnell 1997). After
sub-cluster merging, a new process comes into play: exchange of binary
components (Valtonen \& Mikkola 1991). Thus, the lightest companions
are exchanged by more massive ones and hence, the probability for the
surviving binaries to have nearly equal-mass components is enhanced
(see also Bate, Bonnell \& Bromm 2002b). In addition, where quadruples
are formed involving binary-binary pairs, the surrounding
circum-quadruple disc will tend to drive the total masses of each
binary to similar values.

 A large fraction of the bodies ejected from the unstable
sub-clusters become sub-stellar objects. This is so because the ejected
objects are typically the low mass components of the system and
because, after being ejected from the dense region in which the
multiple system sits, the ejected body is largely deprived of further
accretion (Reipurth \& Clarke 2001; Bate, Bonnell \& Bromm
2002a). Hence, taking into account that all the objects in these
simulations start with a mass close to the opacity limit (a few
Jupiter masses), it is not surprising that many of the ejectae become
brown dwarfs.

The spatial distribution of stars and brown dwarfs within each
multiple (henceforth the {\it configuration}) is shown in Table~1. For
each simulation, the {\it configuration} column contains two rows: the
upper one refers to the internal structure of the multiples at $t =
0.5$ Myr, i.e. at the end of the hydrodynamic calculation (hereafter
{\it hydrodynamic configuration}); and the bottom row corresponds to
$t = 10.5$ Myr, i.e at the end of the $N$-body integration. All the
other columns, except the first two, also refer to both the
hydrodynamic and N-body results.

\subsubsection{Hydrodynamic results}

It can be seen from the {\it hydrodynamic configurations} that the
basic building blocks of the multiples' internal hierarchy are binary
stars. This property reflects the origin of these bound systems as
byproducts of a hierarchical formation process, in which the typical
outcome of a small-$N$ cluster disintegration is a tightly bound
binary star. Subsequent merging of several small-$N$ clusters bind two
or more binaries into one single multiple system. Column 3 ([N$_{\rm
m}$ ; N$_{\rm s}$]) shows on the left the number of stars and brown
dwarfs that remain in multiple systems at the end of the simulations;
in the case that more than one mutually unbound multiple system is
formed, the membership number of each is separated by commas within
brackets. On the right is shown the number of stellar objects that
escaped from the cloud. Overall, the ratio of unbound to bound objects
is approximately 2/3 at 0.5 Myr.

At an age of 0.5 Myr, 60\% of the stars and brown dwarfs are locked in
12 multiple systems, with about a third of the components being
low-mass, weakly bound outliers. Excluding these outliers and unbound
singles, 7\% of the remaining objects are in pure binaries (2
systems), 14\% are in quadruples (2 systems), 35\% are in quintuples
(4 systems), 32\% are in sextuples (3 systems) and 12\% are in
multiples with seven components (1 system).

 Figure~1 shows the value of the mass ratio $q$ (left panel) and
semi-major axis $a$ (right panel) of the resulting multiples as a
function of primary mass (see figure caption for an explanation of the
symbols). It can be seen that no binary star with primary mass
larger than $\approx 0.2$~M$_\odot$ is formed, despite the initial
mass of the components being $200 \times$ smaller. In addition, all
except one binary have $q$ larger than 0.5. The exception appears in
the region located between the solid and dashed lines, where binaries
with brown dwarf companions are found. Apart from this binary, brown
dwarf companions are only found as wide components of high-order
multiples.

\begin{figure*}
\begin{center}
\caption{ Pictorial sketches of the spatial distribution of the
components of a representative sample of the multiple systems
resulting from our simulations. Large dots represent stars in the mass
range $0.1-0.6$~M$_\odot$ and small dots represent {\it outliers}
(M~$< 0.1$~M$_\odot$). The number beside each dashed straight line
gives the distance (in AU) between the sub-systems joined by that
line. These are typical configurations at an age of 0.5 Myr.  On the
bottom right a {\it planetary multiple} is shown; the other systems
involve two or more mutually bound binaries.}
\centerline{\epsfig{file=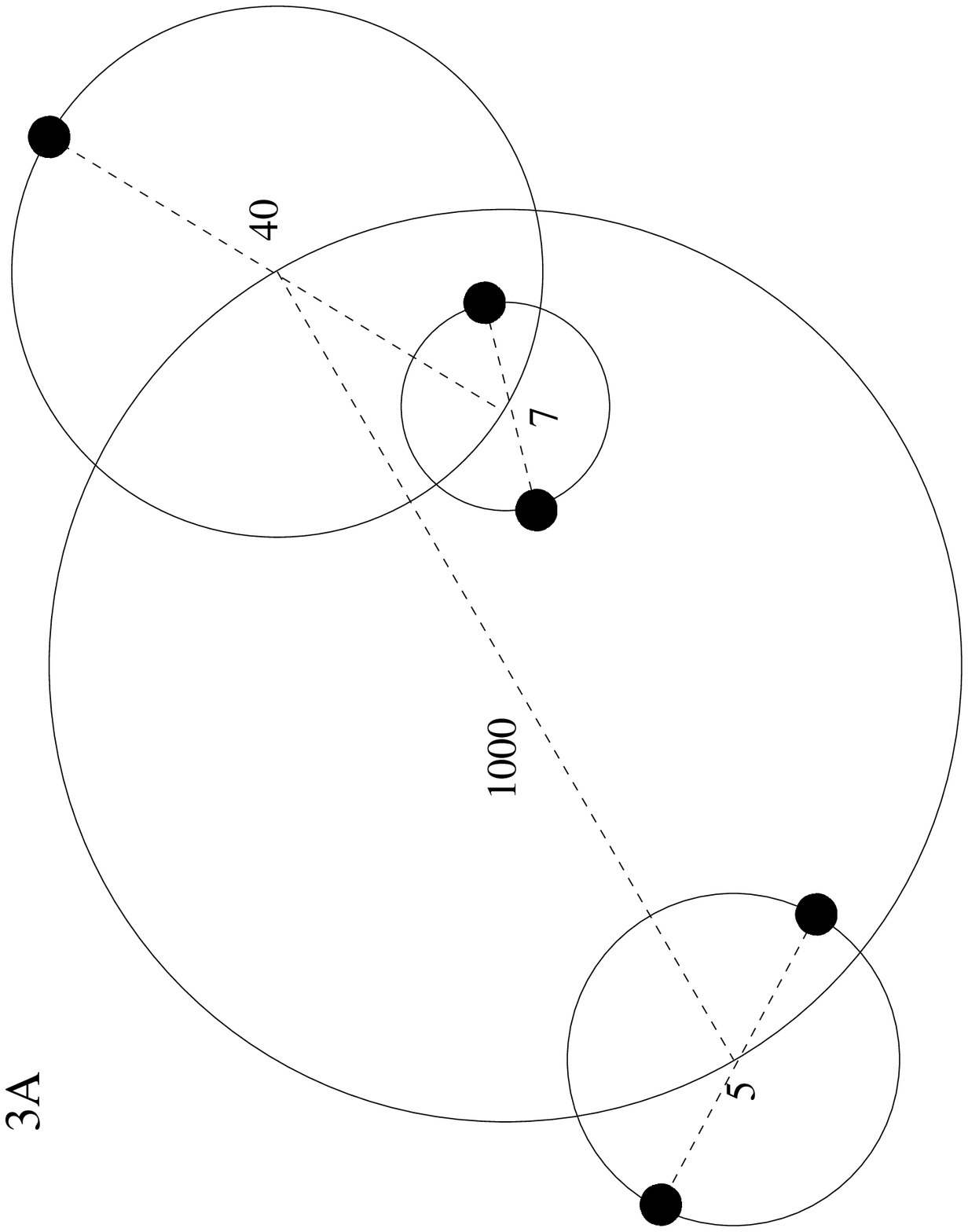,height=7.5cm,angle=270}\epsfig{file=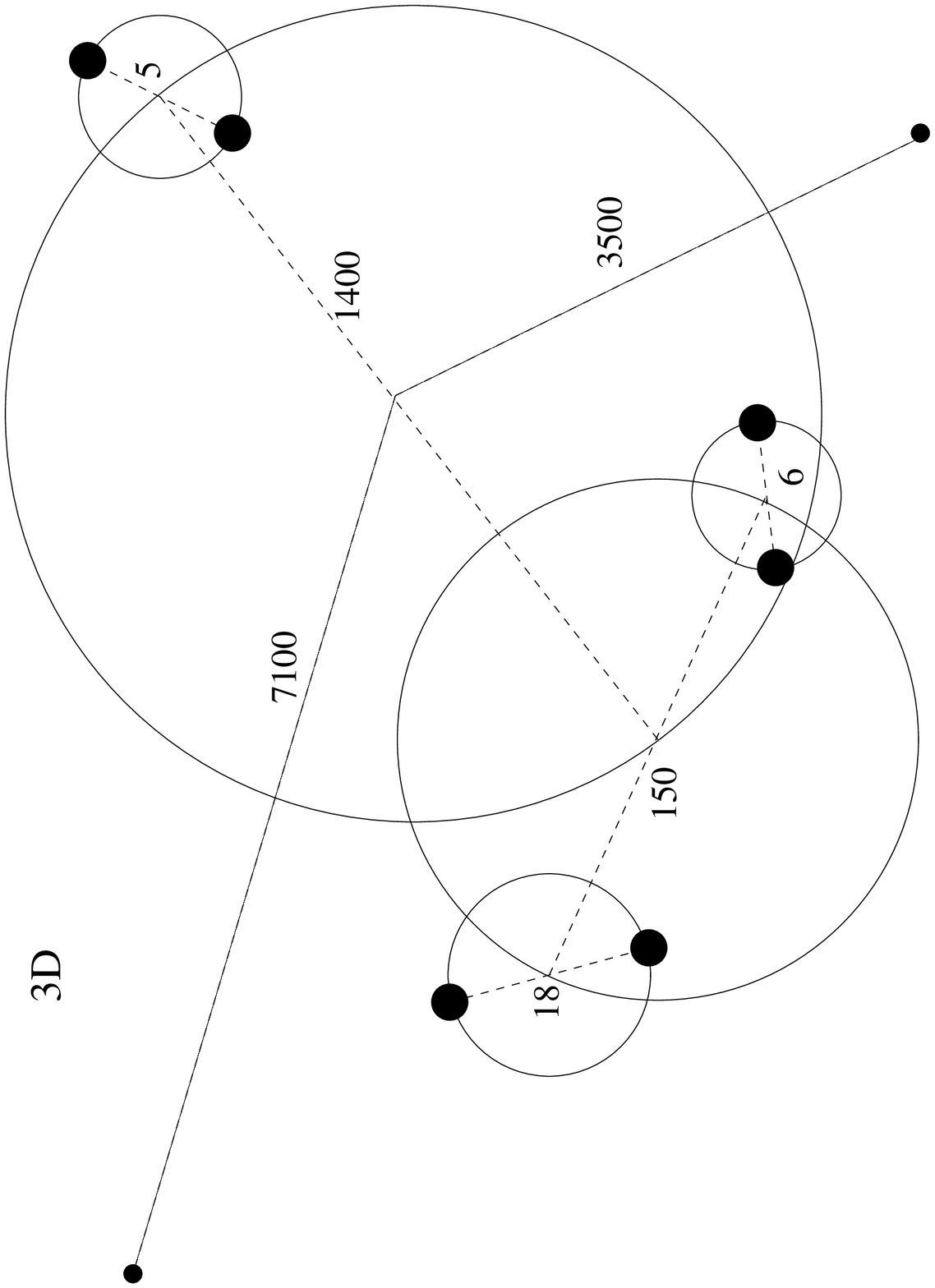,height=7.5cm,angle=270}}
\centerline{\epsfig{file=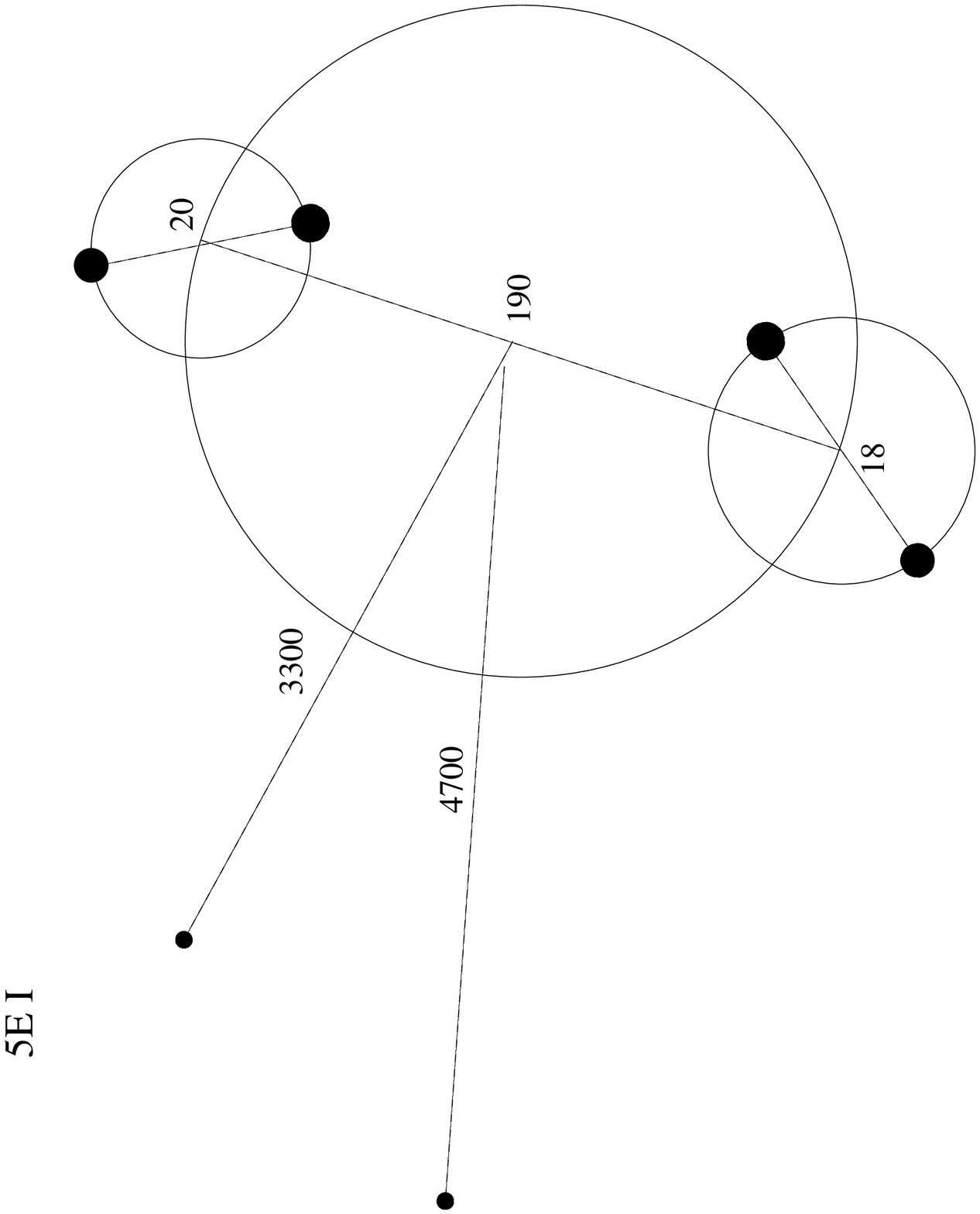,height=7.5cm,angle=270}\epsfig{file=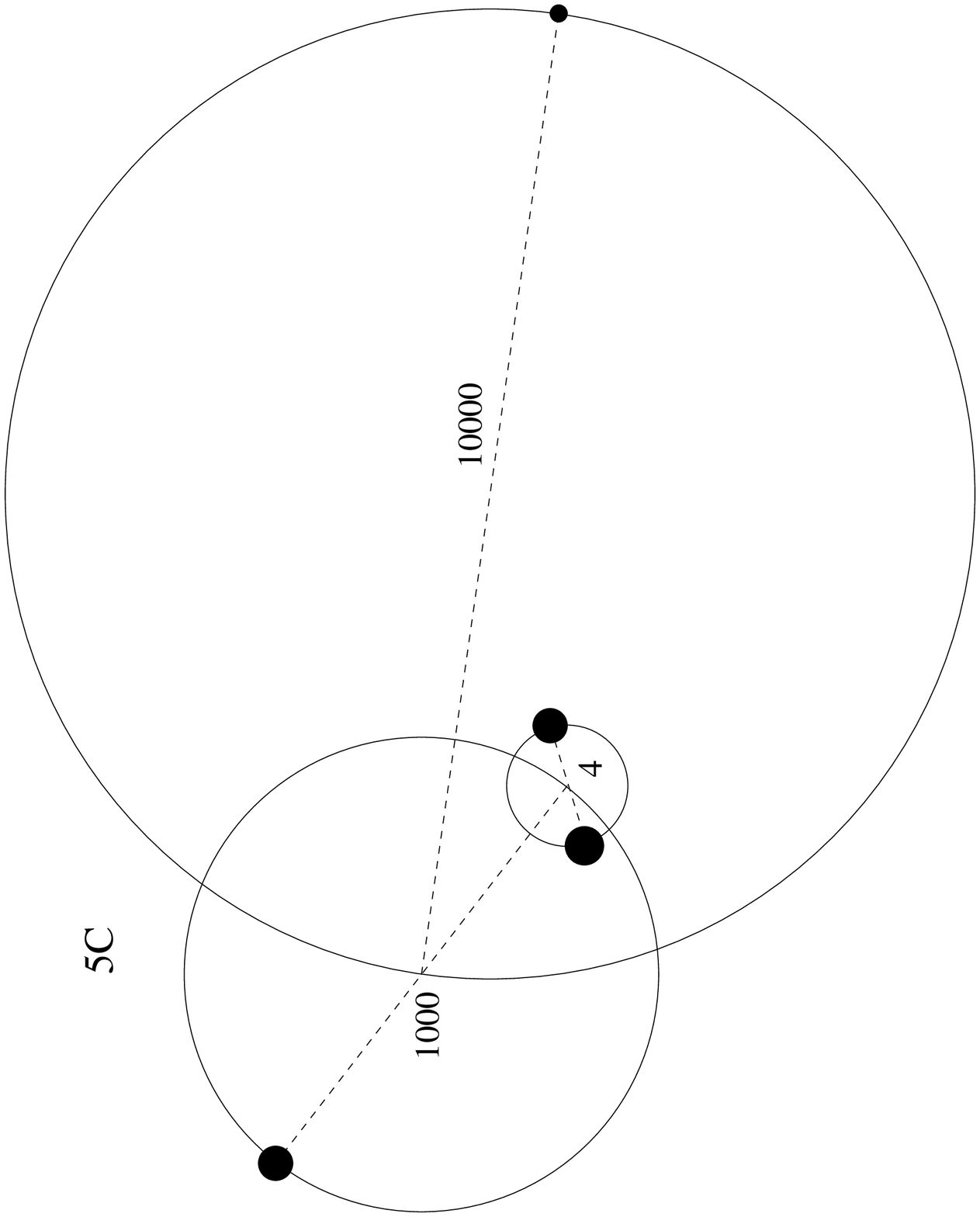,height=7.5cm,angle=270}}
\end{center}
\end{figure*}

 The right panel illustrates two trends. Firstly, high-$N$
multiples are typically {\it wider} than low-$N$ multiples. This trend
reflects the hierarchical or nearly hierarchical spatial distribution
that the multiple's components adopt for the system to attain
dynamical stability -- note that this trend is better visualised if
the symbols rather than the primary masses are considered. The second
trend reflects a shortcoming of these simulations: binaries have a
mean $a$ of $\approx 10-20$~AU, i.e. the vast majority are close
binaries and only 1 system has a $a > 50$~AU. This result is
independent of primary mass. In other words, wide {\it pure} binaries
of any mass are underproduced by our simulations and although closer
binaries might have been formed if the softening length used to smooth
the gravitational force at short distances had been chosen smaller, no
wider system can be produced with these initial conditions. The reason
for this is twofold: first, there is not enough {\it global} angular
momentum initially to form many wide binaries or a very wide binary,
and second but most importantly, the high density of the sub-clusters
prevents wide systems from surviving. This conclusion can be applied
to all the systems underproduced by our simulations: low mass, low
mass ratio and wide binaries share the property of having low binding
energies. Only the systems with large binding energies can survive the
dynamical encounters taking place in the dense clusters formed by
collapsing turbulent flows.

 Figure~2 shows pictorial sketches of the spatial distribution of
the components of a representative sample of the multiple systems
resulting from our simulations, i.e. a binary star orbiting a triple
(top left), a binary quadruple (a binary orbiting another binary) with
some {\it outliers} (bottom left), a binary orbiting a binary
quadruple plus some {\it outliers} (top right), and a {\it planetary}
quadruple (bottom right). The first three classes of multiples
comprise about 75\% of all the multiples formed in the
simulations. The remaining 25\% (the fourth class) is comprised by
{\it planetary} multiples, systems in which companions are not members
of binary/triple systems other than the multiple itself. One of the
noteworthy features of the mass distribution within the multiples is
how similar the masses of all components are, either taking an
individual star or a binary star as the fundamental unit. In other
words, most binaries have very high $q$ and, when two binaries are
bound to each other, the ratio of the total mass of one to the total
mass of the other is also close to unity (in the range $0.5-1$; see
left panel of Figure~1). This pattern of mass distribution can be
readily related to the hierarchical formation mechanism that produces
the multiples: at each sub-cluster merging event, the binary stars
involved interact strongly, exchanging components, and accreting
material with higher specific angular momentum than that of their
orbit. These two processes invariably favour the formation of bound
pairs (a pair of stars in the first small-$N$ cluster, a pair a
binaries after a merging event, etc.) with high values of $q$ ($>
0.5$).

Whereas the first star formation episodes involve as much direct
fragmentation in filaments as disc fragmentation (BBB), at later
times, star formation is dominated by instabilities in
circum-binary/quadruple discs. Most objects formed at late times are
unlikely to remain bound since they start off with a mass close to the
opacity limit for fragmentation whereas the other sub-cluster members
have been accreting for some time and have much larger masses,
therefore being better positioned for future three-body
encounters. The net result is that only those objects formed during
the first free-fall times after star formation begins are likely to
remain in bound structures, whilst the rest will be ejected at very
large separations ({\it outliers}) or simply escape from the cloud.

This is reflected in Table 1 in the penultimate column (N$_{\rm
out}$), which gives the number of objects that being bound to a given
multiple system, have a mass much lower than the other components, and
highly eccentric orbits. Outliers orbit the CoM of the multiple at
large distances, $\approx 10^3 - 10^4$ AU (see Figure~1 [right panel;
crosses correspond to outliers] or the last column of Table~1). The
population of {\it outliers} is mostly comprised by brown dwarfs, and
makes up about 20\% of all the objects in bound systems.

\subsubsection{$N$-body results}

The $N$-body evolution of the stellar systems result in the break-up
of many of the higher-order multiples. At an age of 10.5 Myr, 36\% of
all stars and brown dwarfs are locked in 18 multiple systems, with
only about 2\% of the components being low-mass outliers. Excluding
these, 42\% of the remaining bound objects are in pure binaries (11
systems), 12\% are in triples (2 systems), 15\% are in quadruples (2
systems), 19\% are in quintuples (2 systems) and 12\% are in sextuples
(1 system). The companion frequency $f_{\rm c}$ or average number of
companions per star system is defined as $\frac {b + 2t + 3q + 4k +
...}  {s + b + t + q + k + ...}$, where $s$ represents the number of
singles, $b$ the number of binaries, $t$ the number of triples, $q$
the number of quadruples and $k$ the number of quintuples. This
quantity changes substantially during the 10 Myr of $N$-body
evolution, from 1.0 to 0.3. On the other hand, the multiplicity
frequency $f_{\rm m}$, defined as $\frac {b + t + q + k + ...} {s + b
+ t + q + k + ...}$, shows just a small variation, from 0.18 to
0.17. The multiplicity frequency shows such small variation because
the number of multiples and the number of singles increase during the
$N$-body evolution by similar factors, $\approx 1.4$ and $\approx 1.5$
respectively. Since these factors appear one in the numerator and the
other in the denominator of $f_{\rm m}$, they cancel out almost
completely and thus $f_{\rm m}$ remains practically unchanged. These
two quantities, $f_{\rm m}$ and predominantly $f_{\rm c}$, reflect the
process of multiple disintegration that has taken place during the
$N$-body evolution: a substantial number of high-order multiples have
broken-up into their fundamental units, i.e. pure binaries, which now
comprise the largest group. The fraction of multiples out of the total
number of systems remains almost constant, whereas the average number
of members per system decreases by a factor of 3. This decrement is
predominantly due to the ejection of outliers (only 3 survive after 10
Myr), in other words, the transference of bound objects to the
population of singles. This process is responsible for $\approx 80\%$
of the variation in $f_{\rm c}$. Ejections of outliers have a side
effect: they harden the inner multiple, which very often is a binary
quadruple. After several ejections, the separation between the
binaries may have diminished enough so as to render the multiple
unstable. After several orbital timescales, close binary-binary
interactions result in the disintegration of the multiple, either into
two separate binaries or a binary and two singles. This process,
however, only accounts for $\approx 20\%$ of the variation in $f_{\rm
c}$, which is clearly dominated by the outlier-to-single transference.
A noteworthy feature of this disintegration process is that the bottom
level of the multiple's hierarchical distribution, namely the binary
stars, remains substantially intact. For example, in simulation
$\alpha3C$ two binaries and one triple are mutually bound at 0.5
Myr. After $N$-body integration to 10.5 Myr, these systems have become
mutually unbound, but their internal configuration is still the same.
There is only one case ($\alpha3D$) in which the disintegration of the
multiple results in the exchange of components between two binaries
together with the disruption of one of them.

The typical orbital timescale of the outliers is $\sim 10^6$ yr. After
a few orbits these objects should experience several close encounters
with the inner multiple (since they have very eccentric orbits), thus
making ejection likely to happen during the next few $\times 10^6$ yr
after the start of the $N$-body integration. It is not surprising then
that at an age of 10.5 Myr only three out of the original 30 outliers
have remained in bound orbits, and a substantial fraction of the
multiples have disintegrated. Overall, the ratio of unbound to bound
objects is approximately 5/3 at 10.5 Myr. In addition, the number of
single brown dwarfs has increased by $\approx 60\%$ in the same
period.

We find that the ratio of binary to higher-order multiples at 0.5 Myr
is 2/11, but it rises to 10/8 at 10.5 Myr.  DM91 find a ratio
binary:higher-order multiple of $\approx 5:1$ for the solar
neighbourhood. However, their sample might suffer from a bias against
$N > 2$ systems; e.g. some of the systems might be triples disguised
as binaries either because one of the components might be itself an
unresolved binary or because a faint tertiary companion might remain
undetected. DM91 did spectroscopic follow-up of stars but the wider
systems were drawn from the literature; there could be many missing
wide components, which in turn might themselves be multiples. In fact,
Mayor \& Mazeh (1987) and Mazeh (1990) suggest that 25-50\% of the
spectroscopic binaries might be higher-order multiples. The evidence
for this is deduced from the study of some tight binaries which
exhibit non-negligible residuals after their orbit is fitted by a
Keplerian 2-body orbit. Moreover, spectroscopic studies by Tokovinin
and coworkers (Tokovinin 1997; Tokovinin \& Smekhov 2002) suggest that
a substantial fraction of the components of binary or triple systems
might be themselves unresolved binaries. Therefore, it seems unclear
at the moment whether our results are in agreement with the observed
ratio of binary:higher-order multiples, as deduced from the studies on
individual systems. However, as we will show later, our results
compare well with the photometric width of the main sequence of young
clusters like Praesepe. Therefore, the high incidence of $N>2$
multiples that we predict, given the present observational
constraints, cannot be ruled out.

At the end of the $N$-body integration 17 out of the 18 multiples have
stable configurations (for at least several $\times 10^8$ yr),
according to the criteria of Eggleton \& Kiseleva (1995), whilst the
$18^{th}$ may further decay on a timescale of a few Myr. However,
external perturbations may also affect the stability of these
systems. It is expected that a substantial fraction of all multiple
systems ever formed out of a cloud core will become members of a
larger structure, namely a young stellar cluster or association, or
perhaps an open cluster in the long term (Clarke et al. 2000, Lada \&
Lada 2003). Dynamical interactions within these larger stellar
aggregates may produce the disruption of some of the widest systems
that survived the $N$-body integration (e.g. as in the Orion Nebula
Cluster; Scally, Clarke \& McCaughrean 1999). In a longer term still,
for those systems joining the field population, the gravitational
field of the galaxy will act to disrupt wide systems (Weinberg,
Shapiro \& Wasserman 1987), limiting the maximum separation between
components to a value somewhere around 10$^4$ AU for solar-mass stars
(Close et al. 2003).

 Single and binary stars attain comparable velocities in the range
$1-10 {\rm km~s^{-1}}$. Higher-order multiples display lower velocity
dispersions. This kinematic segregation as a function of $N$ is the
expected outcome of the break-up of unstable multiples, whereby the
ejected objects (typically singles, or less often binaries) acquire
large velocities whereas the remaining more massive multiple recoils
with a lower speed. Therefore, we would expect low mass SFRs like
Taurus or Ophiuchus to display an overabundance of $N > 2$ systems as
lower-$N$ systems may escape more easily the potential well of these
associations. A more detailed discussion on the kinematic properties
of the stellar objects resulting from our simulations is given by
DCB04.

\subsection{Multiplicity fractions}

Overall, the multiplicity fraction $f_{\rm m}$ derived from these
simulations is low: it takes a value close to 0.2, much lower than
that observed in clusters and the field. However, the variation of the
multiplicity fraction with primary mass can be very steep, as Sterzik
\& Durisen (2003) have shown, both observationally and theoretically,
and a low total value may thus mask high and low values in different
mass ranges. Figure~3 shows the dependence of $f_{\rm m}$ on primary
mass, for the hydrodynamic simulations exclusively, at 60\% efficiency
(squares joined by a solid line) and 30\% efficiency (circles joined
by a dashed line). We pointed out in Section~3.1 that the overall
value of $f_{\rm m}$ at 0.5 Myr is very similar to that found at 10.5
Myr. This conclusion also applies to the individual values of $f_{\rm
m}$ in each mass bin. Therefore, the 0.5 Myr 60\% efficiency curve is
also a good representation of the $f_{\rm m}$ dependence on primary
mass 10 Myr later. Also shown in Figure~4 are the observational
results from DM91, Fischer \& Marcy (1992; FM92), Marchal et
al. (2002, M\&02) and Bouy et al. (2003), Close et al. (2003), Gizis
et al. (2003) and Mart\'{\i}n et al. (2003) [BCGM03]. As expected, the
multiplicity fraction is an increasing function of primary mass, with
the most massive stars produced in these simulations having a
multiplicity fraction very close to 1. At the other end, brown dwarfs
are rarely binary primaries. Nevertheless, the shape of the $f_{\rm
m}$ curve is sensitive to the efficiency assumed. For 30\% efficiency,
the results show better agreement with observations at the sub-stellar
regime, while the 60\% efficiency results match more closely the
observed multiplicity fractions for KM stars.

\begin{figure}
\begin{center}
\caption{Multiplicity fraction $f_{\rm m}$ as a function of primary
mass (in M$_\odot$). Dotted vertical lines separate the different mass
bins for which f$_{\rm m}$ has been calculated. Squares joined by a
solid line and circles joined by a dashed line denote the values of
f$_{\rm m}$ at 60\% and 30\% efficiency, respectively. The results
shown in this plot correspond to an age of 0.5 Myr (see text).}
\centerline{\epsfig{file=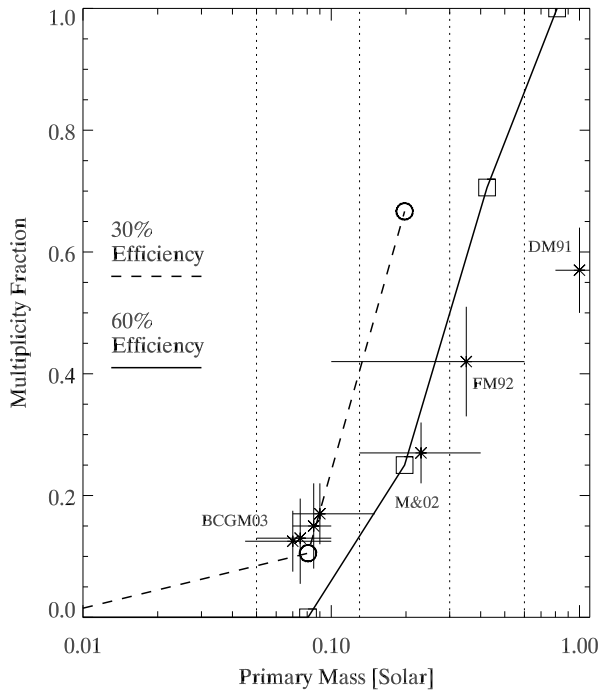,height=10cm}}
\end{center}
\end{figure}

Two points must be stressed however: first, the limitations imposed by
the initial total mass in gas of the cloud, set to 5 M$_\odot$. The
fact that about 10 stars/brown dwarfs form, on average, in each
calculation, and that those with higher mass will be invariably in
bound systems, place a $f_{\rm m}$ value close to 1 in the solar-mass
stars bin. For low-mass star-forming regions (SFR) such as Taurus
(Duch\^ene et al. 1999) or $\rho$ Ophiuchus (Barsony, Koresko \&
Matthews 2003), our choice of 5 M$_\odot$ clouds seem to be
appropriate: both SFRs lack stars more massive than about 2 M$_\odot$
and similarly have a binary fraction approximately twice as high as
that of field stars (which stands at $\approx 50\%$), very close to
what we find. For more massive SFR, models of more massive clouds are
more appropriate: had we modelled a cloud with 50 or 500 M$_\odot$,
the highest value of the multiplicity fraction would have moved to
higher primary masses (e.g. 5 to 10 M$_\odot$; Bonnell \& Bate
2002). This would bring closer agreement with the results in the field
(DM91; Shatsky \& Tokovinin 2002).

Second, the discrepancy with observations at the low mass end (too few
brown dwarf binaries) cannot be alleviated by modelling more massive
clouds. BBB, modelling a 50 M$_\odot$ cloud, found only one brown
dwarf binary candidate out of $\approx$ 20 single brown dwarfs, giving
them a 5\% binary fraction at the sub-stellar regime. In practice,
this binary would have likely been disrupted or simply accreted to
stellar values had the simulation been carried forward longer, since
we do form many of these binaries in our simulations but they soon are
destroyed or become stellar binaries. It remains a puzzle how to
produce a fraction of very low-mass or brown dwarf binaries as high as
is observed. Certainly, clouds with an initial mass of $\sim$ 0.5
M$_\odot$ and a lower number of Jeans masses than used in the present
models, do produce many of these very low-mass binaries. However, in
real clouds, one would expect very low mass cores to interact strongly
with others, as in these simulations, and only the prompt ejection of
a brown dwarf binary via an encounter with a more massive binary could
result in the formation of these systems. However, neither our models
nor those of Bate, Bonnell \& Bromm (2002a) seem to favour this
scenario: ejection of low mass binaries, as opposed to low mass
singles, are rare. One possibility is that the resolution limit
imposed by our choice of the softening radius of the sink particles
may overestimate the number of disruptions taking place among low-mass
binaries. Binaries with separations smaller than $\approx 1$ AU cannot
form, since gravitational accelerations are smoothed within that
radius. Thus, there is a limit to the minimum binding energy that a
bound pair can have. The lower the binding energy of a low-mass bound
pair, the likelier it is its survival to encounters with higher mass
stars. If a brown dwarf binary could become tight enough (i.e. have a
separation $< 1$ AU), it might be able to survive as a bound pair an
encounter and be ejected. This possibility will be explored in future
simulations.

\section{Observational tests}

Some of the properties of our simulated multiple systems can be
immediately compared with current observational data, thus providing
an useful benchmark against which to assess the validity of our
models. Some other properties, however, do not seem to have been
looked upon extensively by observers and may require additional
observational measurements. Therefore, in addition to the conventional
cross-check with observations, the aim of the next subsections is also
to sketch a series of observational tests that, if undertaken, might
help to discriminate among different theoretical models of low-mass
star and brown dwarf formation.

\subsection{Colour-magnitude diagrams}

Colour-magnitude (CM) diagrams provide useful information on the
binarity of a given homogeneous sample of stars and brown dwarfs. Two
different main sequences are frequently discovered in these diagrams,
one corresponding to single stars or individual stars in wide
(i.e. larger than the instrument resolution limit) binary systems, and
the other usually ascribed to unresolved binary stars. If more than
two stars contribute significantly to the total luminosity of a
multiple system, the corresponding point on the CM diagram might be
expected to have a lower magnitude than that of a pure binary of the
same colour, thus broadening the binary sequence. Our simulations
predict the existence of a substantial number (50\% at 10.5 Myr) of
multiple systems made up of three or more stars with comparable
masses. Thus, we would like to know whether a CM diagram of our
simulated stellar systems results in a binary sequence whose width is
consistent with those of real systems or not.

Figure 4 shows the CM diagram corresponding to our multiple systems at
10.5 Myr, after converting masses to $I$ and $K$ magnitudes using the
tracks by Baraffe et al. (1998). We have assumed an age of 600 Myr for
the tracks and a resolution of 200 AU (components of a multiple system
separated by more than 200 AU are considered independent systems when
computing their magnitudes). All systems closer than 200 AU are
stable, thus we do not expect any further evolution of this diagram
for an older system. Red squares correspond to unresolved multiple
systems, and blue open circles to individual stars. Over-plotted, in
solid black circles, are the infrared observational results from the
Praesepe cluster ($\approx$ 600 Myr old and located at $\approx$ 180
pc), by Hodgkin et al. (1999). The stars used in this diagram have
been selected from the Hambly et al. (1995) proper-motion and
photometric survey of Praesepe. The magnitudes of this sample have
been measured from photographic plates, and the resolution limit
ranges between 2~\arcsec and 3~\arcsec, hence our criterion of
2.5~\arcsec ($\approx$ 200 AU at 180 pc) used to identify a given
simulated multiple either as unresolved or as a set of independent
objects.

Three features from Figure~4 may be highlighted: first, a binary
sequence is apparent in the simulated data, {\it and} its {\it width}
is not significantly different to that of the Praesepe cluster, except
for systems redder than $I-K = 2.5$. This seems to suggest that the
formation of triple, quadruple and higher-order multiples of the sort
produced by our models is not ruled out by observations and indeed
might be common in real clusters.

Second, our models fail to produce as many very low-mass binaries as
are observed. The observed binary sequence for Praesepe (as e.g. for
the Pleiades; Pinfield et al. 2003) does continue well into the
sub-stellar regime. As explained in Section 3.2, this observational
result is unaccounted for by our simulations at 60\% star-formation
efficiency, and reproduced only to some extent if very low
efficiencies (probably unphysical for a molecular cloud core) are
assumed.

Third, the sequence of simulated singles does not extend to colours as
blue as those of the multiples. The reason for this is that most of
the mass of the cloud ends up in the inner components of multiple
systems (singles are mostly ejected objects, which remain bound at
large separations or escape completely from the cloud and are
thereafter deprived from further accretion). Had we modelled a more
massive cloud, ejections would have also occurred among higher-mass
objects and the singles sequence might have extended to bluer
colours. Likewise, the lack of wide {\it pure} binaries in our results
contributes to the paucity of objects in the upper part of the singles
sequence. This latter problem might be solved by increasing the net
angular momentum in our models, which initially corresponds to a
$\beta$ parameter (the ratio of the rotational energy to the magnitude
of the gravitational energy of the cloud) of $\sim$ 10$^{-3}$.

\begin{figure}
\begin{center}
\caption{Colour-magnitude (CM) diagram ($I$ vs $I-K$) for the stars
and brown dwarfs formed in the simulations. Red squares represent
unresolved multiples and blue open circles singles. Observational
measurements from the Praesepe cluster are shown as black filled
circles. Resolution at 200 AU (or 2.5~\arcsec at 180 pc)}
\centerline{\epsfig{file=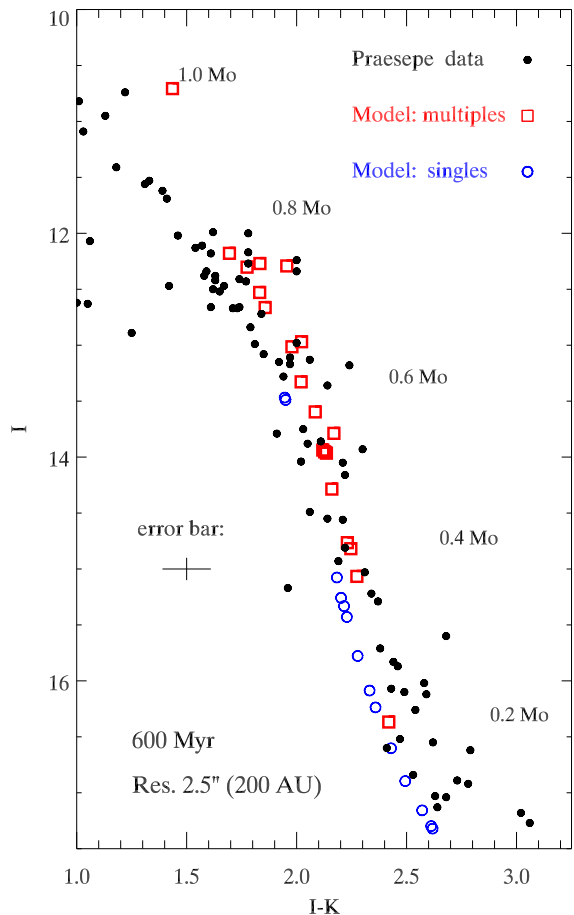,height=13cm}}
\end{center}
\end{figure}

\subsection{High-resolution imaging}

High-resolution imaging with HST or ground-based telescopes using
adaptive optics has proved very successful in resolving tight binary
systems. Most recent studies on binarity among previously unresolved
systems have focused on low-mass stars and brown dwarfs. Bouy et
al. (2003), Close et al. (2003) and Gizis et al. (2003) provide the
most complete set of observations to date of field very low-mass and
brown dwarf binaries. Alternatively, Mart\'{\i}n et al. (2003) have
surveyed young stellar clusters, such as $\alpha$ Persei and the
Pleiades, to pin down the binary fraction among the brown dwarf
population. All these studies suggest that the binary fraction among
very low-mass stars and brown dwarfs is much lower than for G-type
stars, a trend that we also find in our simulations (see Figure
3). Specifically, Close et al. (2003) find a binary fraction in the
range 15\% $\pm$ 7\% for M8.0-L0.5 stars with separations greater than
2.6 AU, and a semi-major axis distribution peak at $\sim$ 4 AU. In
their sample, no very-low mass binary has a separation greater than 15
AU. These results are in agreement with those of Gizis et al. (2003)
and Bouy et al. (2003). Mart\'{\i}n et al (2003) also concludes that a
binary frequency in the range $10\%-15\%$, a bias to separations
smaller than about 15 AU, and a tendency towards high mass ratios ($q
\gsim$ 0.7) are the fundamental properties of brown dwarf binaries. It
must be noted, however, that the $10\%-15\%$ observed binary frequency
is most likely a lower limit and that, in general, there is
considerable uncertainty about the properties of brown dwarf
binaries. In our simulations, most {\it pure} binary systems are tight
(see Figure 1, right panel), with a median semi-major axis of 10 AU,
but few of these systems have primary masses below 0.3 M$_\odot$ 
(for a discussion of this issue, see section~3.2).

Adaptive optics imaging surveys of common associations, such as
Tucana--Horologium (Chauvin et al. 2003), TW~Hydrae (Zuckerman et
al. 2001b), MBM~12 (Hearty et al. 2000) or the $\beta$ Pictoris moving
group (Zuckerman et al. 2001a) are ideal to study the multiplicity
properties of young stars, due to their proximity to Earth ($<$ 70 pc)
and their inferred youth (e.g. $\sim$ 20 Myr for the $\beta$ Pictoris
group). These associations clearly demonstrate the formation of stars
in a clustered mode and, therefore, represent an invaluable laboratory
against which to compare the results obtained by models of clustered
star formation. If dynamical interactions among embedded stars are
indeed as important as shown by our models, this should be reflected
most clearly on the multiplicity properties of the stars populating
these groups: i.e. the binary fraction should be higher than in the
field or in clusters (since many singles should have escaped the
group), low-mass objects should be found to be effectively bound to
more massive binaries/multiples, but at large separations, and
conversely, single unbound brown dwarfs should be rare. Companion
frequencies should also be high (in the range $0.5-1$). Recent results
by Brandeker, Jayawardhana \& Najita (2003) point to this direction:
they find that the multiplicity frequencies in the TW~Hydrae and
MBM~12 groups are high ($0.58 \pm 0.12$ and $0.64 \pm 0.16$,
respectively) in comparison with other young regions such as Taurus or
$\rho$ Ophiuchus (Duch\^ene 1999). The companion frequencies (or
average number of companions per star system) for these two moving
groups are even higher: $0.84 \pm 0.22$ and $0.91 \pm 0.30$,
suggesting that many of the multiples have $N > 2$. These results may
be compared with the value of $f_{\rm c}$ that our models predict if
e.g. 50\% of the singles have left the group at $t= 10.5$ Myr, i.e.,
$f_{\rm c} = 0.7$.

High-resolution imaging has recently provided the first two brown
dwarf candidates orbiting solar-type stars at small separations (Els
et al. 2001; Liu et al. 2002). These brown dwarfs are at distance of
$\approx$ 15-20 AU of the primary. Although these observations imply
that brown dwarf companions do exist at separations comparable to
those of giant planets in our solar system, the frequency of this type
of systems remains unclear: half of the stars harbouring planets have
long term trends in their radial velocities due to unseen companions
(Fischer et al. 2001), but so far no other brown dwarf companion at
close separations has been found. In our simulations we find one
binary (Table~1, ref. $(6-7)$ in $\alpha3$C) out of 24 (at 10.5 Myr)
to have a close brown dwarf companion: this gives a frequency of
$\sim$ 4\%. This binary is composed of a 0.21 M$_\odot$ star and a
0.05 M$_\odot$ brown dwarf ($q = 0.24$), has a semi-major axis of
10~AU and an eccentricity of 0.7.

\subsection{Spectroscopy}

Radial velocity surveys find that the incidence of brown dwarfs within
4 AU of solar-type FGK stars is $<$ 1\%. The evidence for this {\it
brown dwarf desert} at very small separations first emerged from
radial velocity surveys in the late 80's and early 90's (e.g Walker et
al. 1995), and has been confirmed by current high-precision radial
velocity programs (Marcy \& Butler 2000; Halbwachs et al. 2000). Our
results are consistent with this observed {\it brown dwarf desert}: no
binary has a sub-stellar companion closer than 10 AU. The explanation
of this result is very simple: most brown dwarfs companions to stars
are formed via the fragmentation of circumstellar discs, but this same
disc drives quickly the mass ratio of the binary towards unity, due to
the accretion of material with higher specific angular momentum than
that of the binary orbit (Bate \& Bonnell 1997; Bate 2000). In some
other cases, the brown dwarf companion is also likely to be exchanged
by a more massive object after a three-body encounter.

Radial velocity measurements of the components of visual double and
multiple stars (Fekel 1981; Tokovinin 1997; Mazeh et al. 2001;
Tokovinin \& Smekhov 2002) have resulted in the detection of a
substantial fraction of spectroscopic sub-systems. Tokovinin (1997)
points out that the frequency of sub-systems among spectroscopic
binaries with periods shorter than 10 days is at least of 40\%. More
recently, Tokovinin \& Smekhov (2002) have found a frequency of
spectroscopic sub-systems of $\approx$ 20\% per component among 26
resolved visual binaries (i.e. 20\% of these binaries turned out to be
triples or alternatively, 10\% may be binary quadruples). In addition,
18 out of 59 apparently single tertiary components turned out to be
spectroscopic binaries (i.e. 30\% of the systems are quadruples
instead of triples). These results seem to be in agreement with the
high abundance of high-order multiple systems found in our
simulations. Note that although no binaries with semi-major axis
smaller than 1 AU can be produced by our simulations (due to softening
of the gravitational acceleration between point masses at short
distances) there is no reason to believe that such spectroscopic
binaries would not have formed had the resolution limit been much
lower; therefore, a direct comparison of Tokovinin \& Smekhov findings
with our results is indeed meaningful. Further radial velocity
observations of the components of multiple systems would be very
helpful to pin down the real fractions of binary, triple and quadruple
systems, and thus constrain different models of star formation.

\subsection{Very low-mass stars and brown dwarfs at large separations}

In Section 3.1 we pointed out that, in our 0.5 Myr models, very low
mass objects (often brown dwarfs) are very common as companions to
more massive stars, but at large separations. Furthermore, most of
these outliers are bound not to single stars but to close binaries,
triples or binary quadruples. At an age of 10.5 Myr, the fraction of
multiple systems exhibiting brown dwarfs at large separations is much
lower: 3 out of 18, in contrast with the result of 10 out of 13 found
at 0.5 Myr. Nevertheless, 3 out of the 4 brown dwarfs that remain
bound do so orbiting binaries or higher-order multiples at large
separations. The exception is the brown dwarf secondary found in
simulation $\alpha3$C, which is only 10~AU away from the
primary. Therefore, it appears that many bound brown dwarfs ($\sim
3/4$), and most bound brown dwarfs in wide orbits ($\sim 3/3$), should
be orbiting binary, triple or quadruple systems. Currently, 12 brown
dwarf/very low mass companions at wide separations are known
(Kirkpatrick et al. 2001a,b; Wilson et al. 2001), proving that the
{\it brown dwarf desert} does not extend to large separations (Gizis
et al. 2001). Two of these brown dwarfs (Gl337C and Gl584C) are
orbiting visual binaries, at $\approx$ 900 and 3600 AU respectively,
and a third (Gl570D) is bound to a triple system (at a distance of
1500 AU). That is, 3 out of the 11 brown dwarfs orbiting stars at
large separations are known to orbit a binary or triple system. A
clear prediction of our simulations can be applied to the other 8
systems in which a brown dwarf is apparently orbiting a single star. A
large fraction of these singles should turn out, after closer
examination, to be $N \geq 2$ multiples (i.e. a spectroscopic binary,
triple, etc.). This could be tested very simply, since a significant
fraction of all spectroscopic binaries are known to be {\it twins} --
i.e. have nearly equal-mass components (Halbwachs et al. 2003) --, as
is also the case in our models, and twins are easier to detect
spectroscopically than low mass ratio binaries. An observational case
in line with our predictions has been described by Brandeker,
Jayawardhana \& Najita (2003) who have recently shown that the brown
dwarf TWA~5~B is bound at $\approx 120$ AU to TWA~5~A, which in turn
is resolved into a very tight, 3 AU separation, binary (or possibly a
{\it triple}; Mohanty, Jayawardhana \& Barrado y Navascu\'es 2003).

\section{Conclusions}

We have undertaken the first hydrodynamical $+$ $N$-body simulations
of multiple star formation that have produced a statistically
significant number of stable hierarchical systems, with component
separations in the range $\sim 1-1000$ AU. These simulations have
demonstrated that multiple star formation is a major channel for star
formation in turbulent flows. The hydrodynamical simulations are
followed for $\approx 0.5$ Myr; subsequently, the remaining gas is
removed and the stellar systems followed as $N$-body ensembles for an
additional 10 Myr. At this point, all but one of the surviving
multiple systems are stable, according to the criteria of Eggleton \&
Kiseleva (1995).  We find that the properties of the resulting
multiple systems are not significantly sensitive to the large scale
geometry of the cloud -- determined by the turbulence -- but rather to
the dynamical and competitive accretion processes taking place within
the mini-clusters formed out of the collapse and fragmentation of the
cloud.

At an age of $0.5$ Myr, we find that about $60 \%$ of stars and brown
dwarfs are in multiple systems, with about a third of these being low
mass, weakly bound outliers. Excluding these outliers and unbound
objects, $7 \%$ of the remaining objects are in pure binaries ($2$
systems), $14 \%$ are in quadruples ($2$ systems), $35 \%$ are in
quintuples ($4$ systems), $32 \%$ are in sextuples ($3$ systems) and
$12 \%$ are in multiples with seven components ($1$ system). The
companion frequency is therefore very high, $\approx 1$. We find that
our multiples consist of hierarchies of binaries and triples and that
{\it planetary multiples} (in which companions are not members of
binary/triple systems other than the multiple itself) are
comparatively rare (occurring $\sim 25 \%$ of the time). There is a
distinctive pattern of mass distribution within these multiples, with
the mass ratio within binaries, and the mass ratios between binaries,
rarely deviating far from unity (values of $0.5-1$ are typical). On
the other hand, such systems are typically orbited by several low mass
outliers (typically at separations of $\sim 10^4$ AU) on eccentric
orbits. About 90\% of these objects are unstable in timescales of a
few $\times 10^6$ yr (i.e. a few $\times$ their typical orbital
timescale).

We find that the $40 \%$ of objects that are unbound are
overwhelmingly of low mass (median mass $\approx 0.02$
M$_\odot$). Thus our results imply that in the stellar regime, most
stars are in multiples ($\approx 80\%$) and that this multiplicity
fraction $f_{\rm m}$ is an increasing function of mass. In this latter
respect, these results are qualitatively consistent with a large body
of previous works on the decay of small-$N$ systems, both with and
without gas (van Albada 1968; McDonald \& Clarke 1993, 1995; Sterzik
\& Durisen 1998, 2003; DCB03). The high $f_{\rm m}$ values for GK
stars are consistent with adaptive optics measurements of nearby young
associations such as MBM~12 and TW~Hydrae (e.g. Brandeker,
Jayawardhana \& Najita 2003), where multiplicity fractions as high
0.64 are found, and radial velocity surveys of visual binaries
(e.g. Tokovinin \& Smekhov 2002) which raise the percentage of
spectroscopic sub-systems to at least 40\%. Low-mass SFR such as
Taurus or $\rho$ Ophiuchus also show companion frequencies in the
range $0.3-0.5$, comparable to those predicted by our models at later
times. It must be pointed out that the values of the multiplicity
fraction $f_{\rm m}$ for each mass range do not change significantly
during the $N$-body evolution of the systems.

At an age of 10.5 Myr the fraction of bound and unbound objects has
reversed: 40\% remain in multiples and 60\% are singles. The companion
frequency has dropped to $\approx 0.3$ due to the ejection of bound
outliers to the field. This transference of objects from bound to
unbound orbits results in an increase of the number of free floating
brown dwarfs by $\approx 60 \%$. In this 10 Myr time-span, many
multiple systems also experience internal decay: excluding the
remaining 3 outliers, 42\% of the remaining bound objects are in pure
binaries (11 systems), 12\% are in triples (2 systems), 15\% are in
quadruples (2 systems), 19\% are in quintuples (2 systems) and 12\%
are in sextuples (1 system).

We pointed out that low mass stars (and, especially, brown dwarfs) are
{\it locked up} in multiple systems at early times and subsequently
released into the field. This remark needs some qualification
however. Multiple star formation is hierarchical in our simulations,
with structures forming on a particular scale being modified as a
result of subsequent merging with structures on a larger scale.  It is
notable that we find low mass outliers at a separation that is similar
to the initial size of the core, and we speculate that if we had
modelled a larger volume of cloud, rather than isolated cores, we might
have found that these outliers would already have been stripped off by
interactions with structures on a larger scale before some of them
could settle into stable orbits. This suggests that such outliers
should be sought in imaging of relatively isolated pre-main sequence
groups and may explain why no such outliers have been detected through
deep imaging of multiple systems in Taurus (G. Duch\^ene, private
communication).
   
If such outliers do indeed survive the formation process, then about
10\% of them are in stable hierarchical orbits at 10.5 Myr.  We would
thus expect some brown dwarfs to remain in the outer reaches of
multiple systems even in the field.  Our simulations can thus
accommodate the existence of systems with brown dwarfs as wide
companions (Gizis et al. 2001) but {\it only if the primaries of these
systems are themselves multiple systems}. (We find that three out of
the four bound brown dwarfs present at $t=10.5$ Myr are orbiting a
multiple system).  We therefore predict that the primaries of binaries
containing a brown dwarf in wide orbit should themselves be multiple
systems.

We have examined how well the products of our simulations compare with
the properties of real stellar systems as deduced from the colour
magnitude diagram of young clusters (specifically the infrared colour
magnitude diagram for Praesepe). Our simulation data compares very
favourably with the width of the main sequence in the mass range
0.4-1 M$_\odot$; indeed the spread of the main sequence in this mass
range appears to {\it require} that stars are commonly assembled into
high order multiple systems, although the number of outliers from a
pure equal mass binary sequence is not large.

The comparison with the Praesepe colour magnitude diagram however
illustrates two problems with the simulation results.  Firstly, the
simulation produces no single star sequence at masses greater than
0.35 M$_\odot$ (colour bluer than $I-K=2$; there are only two blue
circles above 0.35 M$_\odot$, at 0.581 and 0.584 M$_\odot$), whereas
the observational data shows such a sequence, indicating that single
stars and/or low mass ratio binaries are produced in this colour
range. We could probably alleviate this problem by modelling a more
massive cloud. This would increase the maximum mass of stars produced
and enable some stars in the colour range considered to be ejected as
singles by encounters with more massive stars.  Although this would
bring closer agreement with the observed colour magnitude diagram, the
lack of low mass ratio binaries in our simulations is in conflict with
independent evidence from field binary surveys, such as that of DM91
for G dwarfs.  The DM91 data (containing binaries with median
separation $30$ AU) showed a distribution with mass ratio ($q$) that
rose with decreasing $q$ towards their completeness limit of $q\sim
0.2$, whereas our mass ratio distribution (with the exception of the
very low mass outliers) is strongly concentrated between $0.5$ and
unity.  This inability to reproduce enough extreme mass ratio systems
is a feature of all hydrodynamic modelling of multiple systems to
date. It has not however been widely discussed previously, since BBB
focused on the close systems, where the observed mass ratio
distribution is in any case much more concentrated towards unit mass
ratio than for the binary population as a whole (Mazeh et al. 1992;
Halbwachs et al. 2003) and where the simulations very naturally
reproduce the observed brown dwarf desert at {\it very small}
separations (Marcy \& Butler 2000; Halbwachs et al. 2000).

The second problem revealed by the colour magnitude diagram relates to
the fact that the simulations produce almost no binaries with
primaries redder than $I-K=2.5$ whereas the data exhibits a pronounced
scatter (consistent with large numbers of binaries) in this mass range
($0.15-0.4 M_\odot$).  This scarcity of binaries with low mass
primaries in our simulations also conflicts somewhat with the results
of binary surveys among M dwarfs (Fischer \& Marcy 1992; see Figure 4,
this paper) and brown dwarfs (Close et al. 2003; Gizis et al. 2003;
Mart\'{\i}n et al. 2003; Bouy et al. 2003).

In summary, we have found that our simulations produce large numbers
of hierarchical multiple systems and that relatively isolated young
multiples may harbour weakly bound brown dwarf outliers, as a relic of
the hierarchical formation process in turbulent flows. We predict that
where brown dwarfs are found in wide orbits, the primary should itself
turn out to be a multiple.  The simulations are consistent with a
number of observational constraints: the high (but presently poorly
constrained) incidence of hierarchical multiples among field stars and
pre-main sequence stars, the absence of brown dwarfs as close
companions to normal stars (the brown dwarf desert) and, at a
qualitative level at any rate, the positive dependence of the binary
fraction on primary mass. There are two areas in which the simulations
are not able to replicate observations properly: the simulations
under-produce the binary fraction at low masses (M dwarfs and brown
dwarfs) and also do not generate enough wide stellar pairs with low
mass ratios. 
?
\section*{Acknowledgments}

EJDD is grateful to the EU Research Training Network {\it Young
Stellar Clusters} for support. CJC gratefully acknowledges support
from the Leverhulme trust in the form of a Philip Leverhulme Prize. We
thank an anonymous referee for suggestions that substantially improved
the clarity of the paper. The computations reported here were
performed using the U.K. Astrophysical Fluids Facility (UKAFF).

\label{lastpage}

\end{document}